\documentclass[3p,review,sort&compress]{elsarticle}
\usepackage{graphicx} 
\usepackage{dcolumn} 
\usepackage{amsmath,amssymb,bm} 
\usepackage{hyperref}
\hypersetup{colorlinks=true,citecolor=blue,linkcolor=red,urlcolor=blue}
\usepackage{xcolor}
\usepackage{soul}
\usepackage{titletoc}
\begin{document}

\begin{frontmatter}
\title{Electronic and phonon contributions to the Thermoelectric properties of newly discovered half-Heusler alloys XHfPb (X= Ni, Pd, and Pt)}
\author[1]{Paul O. Adebambo}\ead{adebambo@physics.unaab.edu.ng}
\author[1]{Gboyega A. Adebayo}\ead{adebayo@physics.unaab.edu.ng}
\author[2]{Roberto Guerra}\ead{roberto.guerra@unimi.it}
\author[3]{Davide Ceresoli\corref{cor1}}\ead{davide.ceresoli@cnr.it}
\cortext[cor1]{Corresponding author}

\address[1]{Department of Physics, Federal University of Agriculture, PMB 2240, Abeokuta, Nigeria}
\address[2]{Center for Complexity and Biosystems, Department of Physics, University of Milan, via Celoria 16, 20133 Milano, Italy}
\address[3]{Consiglio Nazionale delle Ricerche, Istituto di Scienze e Tecnologie Chimiche ``G. Natta'', (CNR-SCITEC), via Golgi 19, 20133 Milano, Italy.}

\date{\today}

\begin{abstract}
In this work we calculate the thermoelectric figure of merit of XHfPb (X= Ni, Pd, and Pt) by computing the both the power factor and the lattice thermal conductivity by first principles. We make reasonable approximations: we use the Constant Relaxation Time Approximation (CRTA) to compute the electron transport contribution and the modified Debye-Callaway model to calculate the thermal lattice conductivity. We also report the dielectric properties of these semiconductors and the mode Gr\"uneisen parameters. Not surprisingly we find that the average Gr\"uneisen coefficient correlates with the tehrmal conductivity. Next, we consider a realistic relaxation time $\tau$ and carrier concentration $n$ from experimental data on ZrHfPb and obtain the figure of merit $ZT$ as a function of temperature. Our main finding is that despite the Pt is isoelectronic with Ni and Pd, the $ZT$ of PtHfPb is larger and behaves differently from the other two materials, suggesting that PtHfPb is better suited for high temperature thermoelectric generators.
\end{abstract}

\end{frontmatter}

\section{Introduction}\label{sec:intro}
With their existence known for more than a century, the Heusler compounds gained increased interest over the last decades because of their broad range of exciting properties and potential applications~\cite{Kanomata2009}. In particular, Heusler alloys have been identified as possible candidates for high-temperature structural applications and as thermoelectric (TE) materials~\cite{Quinn2021}. Structurally, Heusler alloy exists in two forms: Full-Heusler (FH) alloys and Half-Heusler (HH) alloys. The FH alloys occupy the space group {\color{red}$F$$m$-3$m$} (\#225) with general formula A$_2$BC and crystallize in the L2$_1$ structure. At the same time, the Half-Heusler alloys have the space group number {\color{red}{$F$$\bar4$3$m$}} (\#216) with general formula ABC and crystallize in C1$_b$ structure. In both cases, A and B atoms are transition metals, while the C atoms can be $sp$ metals or semiconductor elements belonging to groups III, IV, V of the periodic table~\cite{Galanakis2002}. Among the different classes of Heusler alloys, non-magnetic HH alloys with 18 valence electrons are particularly suitable to TE applications because they have narrow band gaps, high Seebeck coefficients, and large number of valence electrons per unit cell~\cite{Yang2008,Schwall2011,Chen2013}.

The efficiency of TE effect in materials is quantified by the dimensionless figure of merit, $ZT = \sigma S^2{T}/\kappa,$ which contains the electrical conductivity $\sigma$, the Seebeck coefficient $S$, temperature $T$, and the thermal conductivity $\kappa$~\cite{Kieven2010,Ding2015}. Clearly, large TE efficiencies simultaneously require a large electrical conductivity and a small thermal conductivity in a wide temperature range. To date one of the best performing TE, Bi$_2$Te$_3$, is found to operate optimally in the range between $300$ and $500$\,K~\cite{Snyder2008}. For high temperature energy harvesting, PbTe and GeTe have been suggested as possible alternatives to Bi$_2$Te$_3$~\cite{Gaultois2013,Yee2013}.

Besides, several investigations on non-magnetic HH alloys have also been carried out, showing high TE performance and opening to possible industrial applications~\cite{Andrea2015,Aliev1990,Aliev1991, Kawaharada2004,Sakurada2005,Zhu2015,Rausch2015,Tang2018,Miyazaki2020,Eliassen2017,Carrete2014,Larson1999,Hohl1999}. Recently, Wang et al.\ studied the electronic structure, and the TE properties of NiHfPb and PdHfPb along with some other HH alloys, providing reference values for the Seebeck coefficient ($S$), the power factors, and $ZT$ of these HH alloys~\cite{Wang2016}. Kaur et al.\ studied the TE properties of PtHfPb HH alloy, reporting $ZT$, $S$, $\sigma$, and the electron thermal conductivity as a function of temperature~\cite{Kaur2017}. Despite the vast information available on some of these materials, very little attention has been given to the dielectric properties and to the anharmonic properties (i.e.\ the Gr\"uneisen parameters) which affect the lattice thermal conductivity, and, in turn, the figure of merit $ZT$ of these alloys.

In the present study we focus on the newly discovered~\cite{Gautier2015,Yan2022} HH alloys XHfPb (X= Ni, Pd, and Pt) as potential candidates for TE applications. We start from density functional theory (DFT) total energy and lattice dynamics calculations, and then we combine semi-classical Boltzmann transport theory, and a modified Debye-Callaway model for the thermal conductivity to obtain the thermoelectric properties and figure of merit, as a function of temperature and carrier concentration. In Sec.~\ref{sec:methods} we discuss the methods and protocols of our calculations, in Sec.~\ref{sec:results} we present and discuss our results, and in Sec.~\ref{sec:conclusion} we summarize our conclusions.

\section{Computational methods}\label{sec:methods}
We carried out DFT non spin-polarized calculations with the Perdew-Burke-Ernzerhof (PBE) exchange-correlation functional~\cite{Perdew1981}, using scalar-relativistic PAW pseudopotentials with non-linear core correction. We set a kinetic energy cut-off of 50\,Ry for the wavefunctions and of 400\,Ry for the charge density. To obtain the ground state and equilibrium volume, we used a reciprocal-space uniform mesh of 8$\times$8$\times$8 k-points. Then, to obtain accurate TE properties, we used a finer 12$\times$12$\times$12 mesh in subsequent non-self-consistent field calculations. All DFT calculations were performed with Quantum ESPRESSO~\cite{Giannozzi2009,Giannozzi2017}.

\begin{figure}[tb!]
    \centering
    \includegraphics[width=0.6\columnwidth]{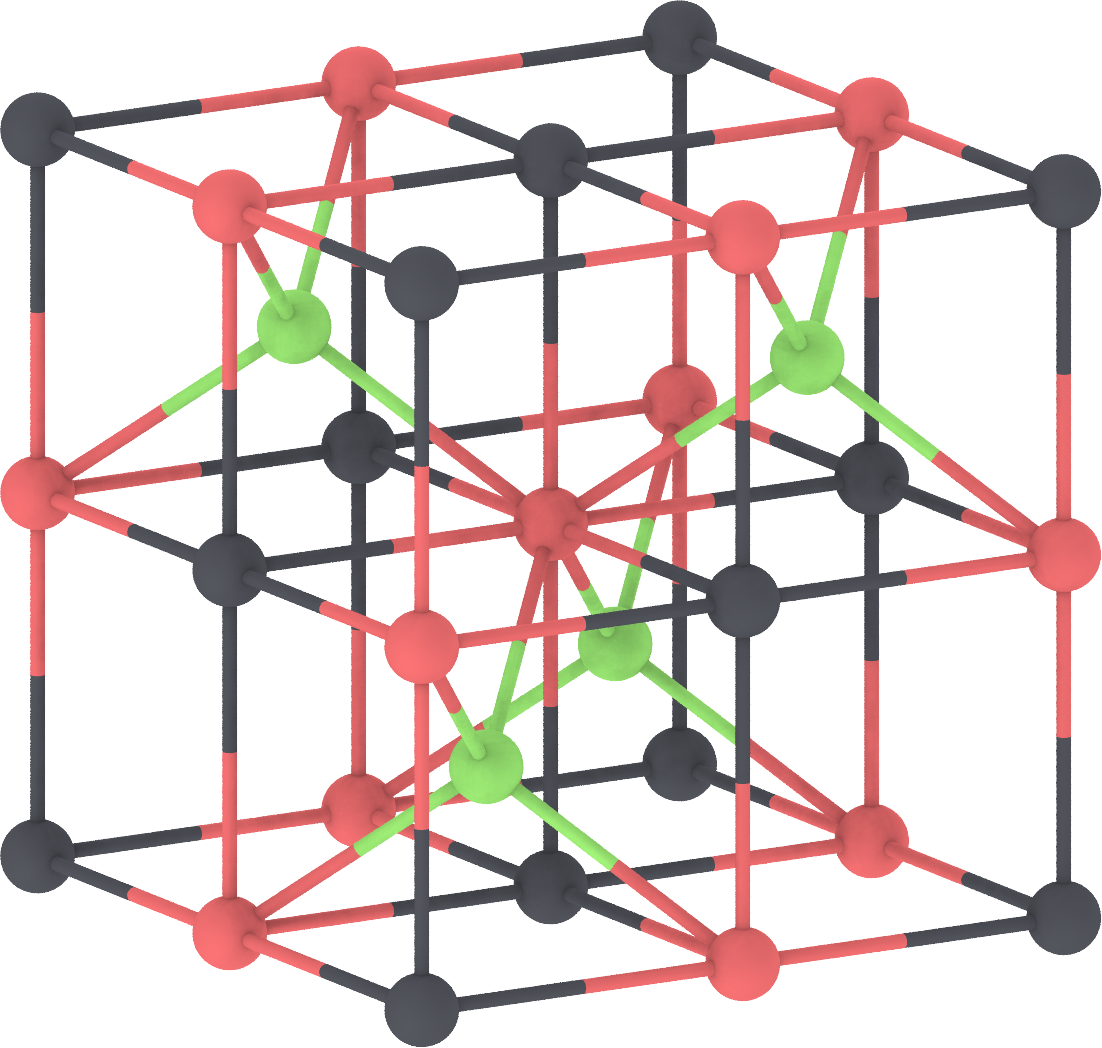}
    \caption{
       Stick-ball representation of the XHfPb Half-Hausler conventional cell. X, Hf, and Pb atoms are colored in green, red, and grey, respectively. Bonds are drawn solely as reference for perspective view.
    }
    \label{fig.structure}
\end{figure}

In all calculations, the XHfPb atoms are arranged such that Hf is placed at {\color{red}4b:($1/2$,$1/2$,$1/2$), X atoms are positioned at 8c:($1/4$,$1/4$,$1/4$), and Pb is located at 4a:($0$,$0$,$0$) in the primitive cell}, as depicted in Fig.~\ref{fig.structure}. {\color{red} Hf and Pb atoms have 4 X first neighbors,
while X atoms have 4 Hf and 4 Pb first neighbors}. The optimal lattice constant and the corresponding bulk modulus have been calculated by fitting the Murnaghan's equation of state~\cite{Murnaghan1944} to the energy-volume curve. The results, reported in Table~\ref{table:EOS}, are in excellent agreement with other calculations~\cite{Wang2016,Kaur2017} and with available experimental data~\cite{Gautier2015}.

\begin{table}[tb!]
    \centering
    \begin{tabular}{c c c c c} 
    Alloy & Ref. & a(\AA) & $B_{0}$(GPa) & $B_{0}^\prime$\\
    \hline
    NiHfPb &Present work& 6.216 &115.0 &4.88\\
           &Calc.\cite{Wang2016}&6.252&\\      
    PdHfPb &Present work& 6.458 & 110.2 &5.01\\
           &Calc.\cite{Wang2016}&6.485&&\\
    PtHfPb &Present work& 6.485 & 124.3 &4.96\\       
           &Expt.\cite{Gautier2015}&6.485&&\\
          &Calc.\cite{Kaur2017}&6.495&113.1&4.85\\   
    \end{tabular}
    \caption{
       Calculated lattice constants, bulk moduli and pressure derivative for HXfPb (X= Ni, Pd, and Pt) compared with other theoretical and available experimental values.}
    \label{table:EOS}
\end{table}

To account for the electronic part of the TE properties, we adopted the semi-classical Boltzmann transport theory as implemented in the BoltzTraP code (version 1.2.5)~\cite{Madsen2006} with the main assumptions being the rigid band and the Constant Relaxation Time Approximation (CRTA). For the evaluation of the electrical conductivity, Seebeck coefficient and electron thermal conductivity in the CRTA we refer the reader to the formulas reported in Ref.~\cite{Madsen2006}. 

To obtain the force constant matrix and the phonon dispersion, we used the finite displacement method~\cite{Kresse1995,Parlinski1997} as implemented in the PHONOPY code~\cite{Togo2008,Togo2015,Akamatsu2014,Skelton2014}, with a 2$\times$2$\times$2 supercell and an atomic displacement of 0.01\,\AA. To obtain the volume Gr\"uneisen parameters, defined as:
\begin{equation}
  \gamma_{\mathbf{q}j} (V) = - \frac{V}{\omega_{\mathbf{q}j(V)}} \frac{\partial \omega_{\mathbf{q}j}(V)}{\partial V},
\end{equation}
we calculated the force constants matrix and the dynamical matrix for a $\pm0.5$\% change of the equilibrium volume.

To calculate the phonon thermal conductivity (the electronic thermal conductivity, usually small, was calculated by BoltzTraP), we used the modified Debye-Callaway model, as implemented in the AICON2 code~\cite{Fan2020,Fan2021}. In this approach the lattice thermal conductivity $\kappa$ is equivalent to the weighted average of the three acoustic and one effective optical branches~\cite{Morelli2002}. The phonon-phonon scattering (normal and Umklapp) is estimated from the Gr\"uneisen parameters, the mean atomic mass in the crystal, phonon velocities and Debye temperature for each phonon branch. The relevant equations are  (28)--(30) of Ref.~\cite{Fan2021}. This approximation works well at high temperature where boundary and impurity scattering become negligible~\cite{Fan2020}. Finally the thermal conductivity is evaluated as a function of temperature $T$ via the linearized Boltzman transport equation (BTE).

The modified Debye-Callaway model is not computationally intensive and showed to be sufficiently accurate to describe the thermal conductivity of a variety of systems even with anisotropic structure like Bi$_2$Se$_3${\color{red}~\cite{Fan2021}}.
This method is well suited for high-throughput quick screening of TE candidate materials and avoids the expensive calculation of third order energy derivatives~\cite{Togo2015,Mizokami2018}, electron-phonon effects on the phonon distribution~\cite{Zhou2021} and methods beyond CRTA~\cite{Jayaraj2022}.

\section{Results and discussion}\label{sec:results}

\subsection{Electronic properties}
\begin{figure}[tb!]
    \centering
    \includegraphics[width=0.6\columnwidth]{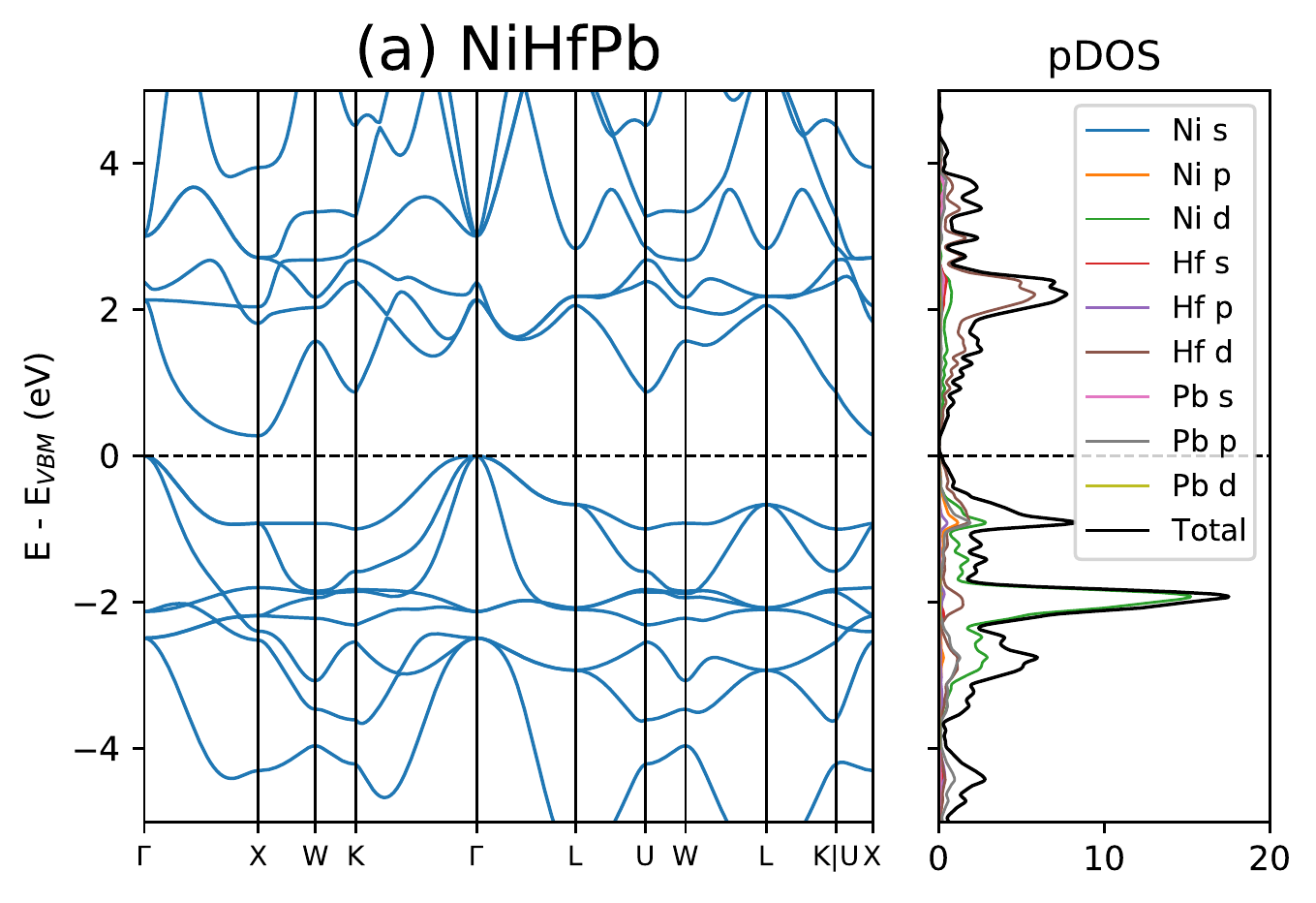}
    \includegraphics[width=0.6\columnwidth]{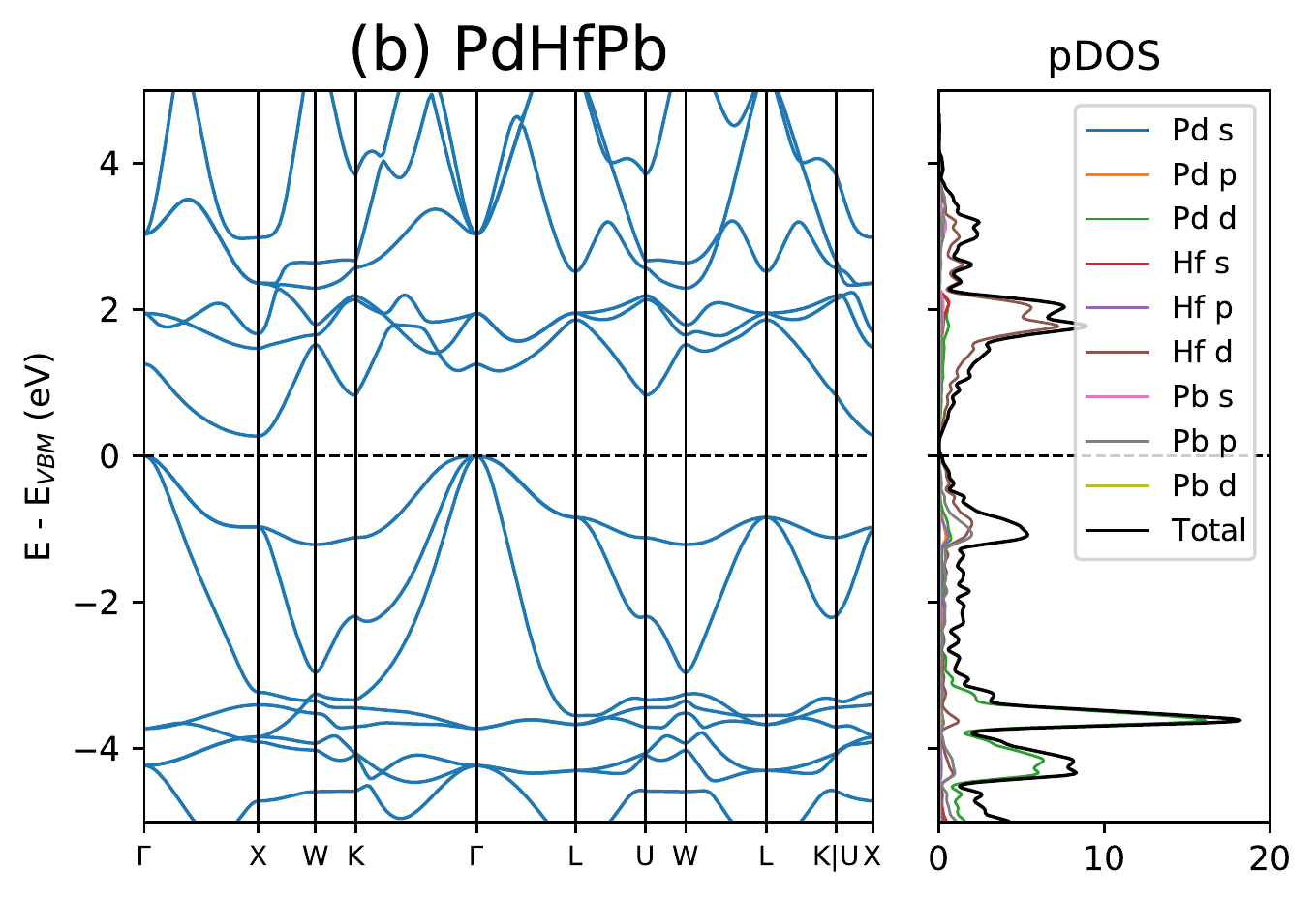} 
    \includegraphics[width=0.6\columnwidth]{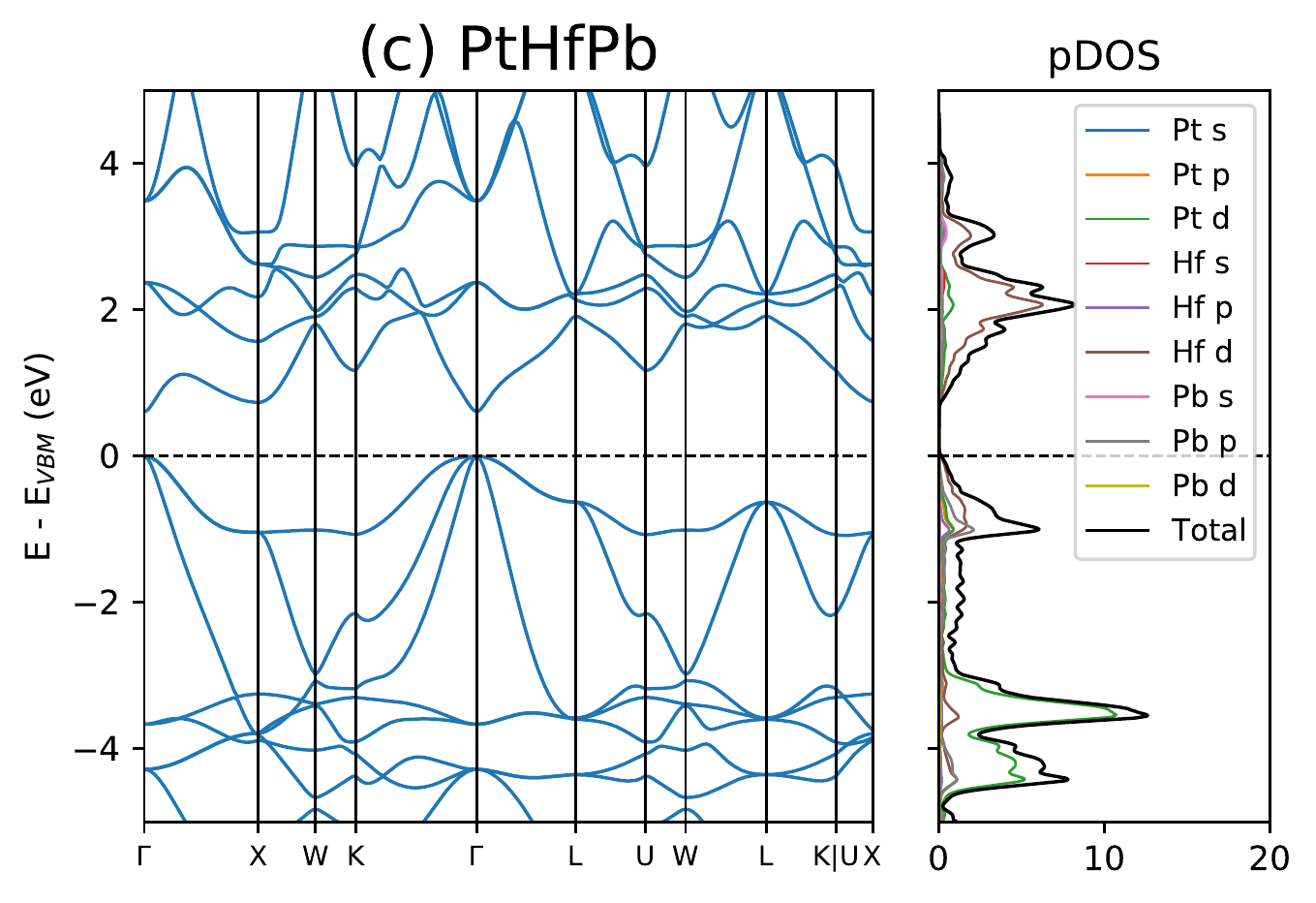}
    \caption{Electronic band structure and projected density of states of (a) NiHfPb, (b) PdHfPb and (c) PtHfPb.}
\label{fig:banddos}
\end{figure}

The electronic band structure and the projected density of state (PDOS) of HXfPb (X=Ni, Pd, and Pt) were calculated along with the high symmetry directions in the Brillouin zone (BZ), as shown in Fig.~\ref{fig:banddos}. The Fermi level was placed at the top of the valence band.
In all three materials the valence band maximum (VBM) is at $\Gamma$, with a curvature appearing to be similar for the three alloys. Conversely, the curvature of the conduction bands at $\Gamma$ changes sign from being negative in NiHfPb and PdHfPb, to positive in PtHfPb. As a consequence, NiHfPb and PdHfPb result to have an indirect band gap with the conduction band minimum located at the X point. Upon substitution of Ni and Pd with Pt, the first conduction band energy increases at X and decreases at $\Gamma$, thus PtHfPb showing a direct band gap.
We estimated the PBE energy gap value of NiHfPb, PdHfPb and PtHfPb alloys to be $0.2769$, $0.2761$ and $0.6070$~eV respectively. {\color{red}To provide a more realistic estimate of the band gaps, we calculated the highest VB and lowest CB eigenvalues with HSE06~\cite{HSE06}. These results are reported in Tab.~\ref{tab:pbehse} and show that the HSE06 conduction band is practically rigidly shifted by $\sim$0.6~eV up in energy with respect to PBE. At the HSE06 level, the band gap values of NiHfPb, PdHfPb and PtHfPb alloys are $0.558$, $0.6630$ and $1.2688$~eV respectively. Given the small electron/hole doping and the almost rigid shifts, we thus expect that moving from PBE to HSE06 would yield a minor effect on the calculated thermoelectric properties.}
The projected density of states (DOS) of the three alloys shows that the VBM and the CBM originates from the Hf $d$ orbital with sizable contributions from the X $d$ orbitals. The latter display sharp peaks at $-2$\,eV for Ni, and at $\sim-4$\,eV for Pd and Pt. 

\begin{table}[]
    \centering
    \begin{tabular}{c | c | c c | c c}
    \hline
    NiHfPt & k point & PBE VB & PBE CB & HSE06 VB & HSE06 CB \\
    \hline
    &$\Gamma$ &  0.0000 & 2.1374 &  0.0000 & 2.8317 \\
    &X        & -0.9165 & 0.2769 & -1.3371 & 0.5580 \\
    &W        & -0.9145 & 1.5754 & -1.3581 & 2.1112\\
    &K        & -0.9967 & 0.8729 & -1.3548 & 1.3460\\
    &L        & -0.6612 & 2.0628 & -0.8749 & 2.7762 \\
    &U        & -0.9967 & 0.8729 & -1.3548 & 1.3460 \\
    \hline
    \end{tabular}
    \begin{tabular}{c | c | c c | c c}
    \hline
    PdHfPt & k point & PBE VB & PBE CB & HSE06 VB & HSE06 CB \\
    \hline
    &$\Gamma$ &  0.0000 & 1.2596 &  0.0000 & 1.9445 \\
    &X        & -0.9655 & 0.2721 & -1.4125 & 0.6630 \\
    &W        & -1.2082 & 1.5288 & -1.5380 & 2.0317 \\
    &K        & -1.1133 & 0.8321 & -1.4465 & 1.3212 \\
    &L        & -0.8356 & 1.8651 & -1.0129 & 2.5013 \\
    &U        & -1.1133 & 0.8321 & -1.4465 & 1.3212 \\
    \hline
    \end{tabular}
    \begin{tabular}{c | c | c c | c c}
    \hline
    PtHfPt & k point & PBE VB & PBE CB & HSE06 VB & HSE06 CB \\
    \hline
    &$\Gamma$ &  0.0000 & 0.6070 &  0.0000 & 1.2688 \\
    &X        & -1.0419 & 0.7352 & -1.4812 & 1.3032 \\
    &W        & -1.0096 & 1.8090 & -1.2718 & 2.4566\\
    &K        & -1.0747 & 1.1698 & -1.3740 & 1.7796\\
    &L        & -0.6265 & 1.9157 & -0.7356 & 2.4938 \\
    &U        & -1.0747 & 1.1698 & -1.3740 & 1.7796 \\
    \hline
    \end{tabular}
    \caption{Highest valence band (VB) and lowest conduction band (CB) eigenvalues calculated with PBE and HSE06 at high symmetry points. The values are in eV and are referenced to the top of the valence band ($\Gamma$) in each case.}
    \label{tab:pbehse}
\end{table}

The calculated band structures and density of states of NiHfPb and PdHfPb are in good agreement with the previous calculations of Wang et al.~\cite{Wang2016}. However the band structure of NiHfPb and PtHfPb differ from the work of Kaur et al.~\cite{Kaur2017} as well as from calculated band structures from the Materials Project~\footnote{Materials Project NiHfPb: \url{https://materialsproject.org/materials/mp-961732/}} and~\footnote{Materials Project PtHfPb: \url{https://materialsproject.org/materials/mp-961715/}}. The reason is that in those structures Hf has been placed wrongly in the $4c$ Wyckoff site (1/4,1/4,1/4) and Pb in the $4a$ site (0,0,0). We verified that our XHfPb structure, described in Sec.~\ref{sec:methods} is the correct one and has the lowest energy among the possible other atomic arrangements.

\subsection{Dynamical and dielectric properties}
\begin{figure}[tb!]
    \centering
    \includegraphics[width=0.6\columnwidth]{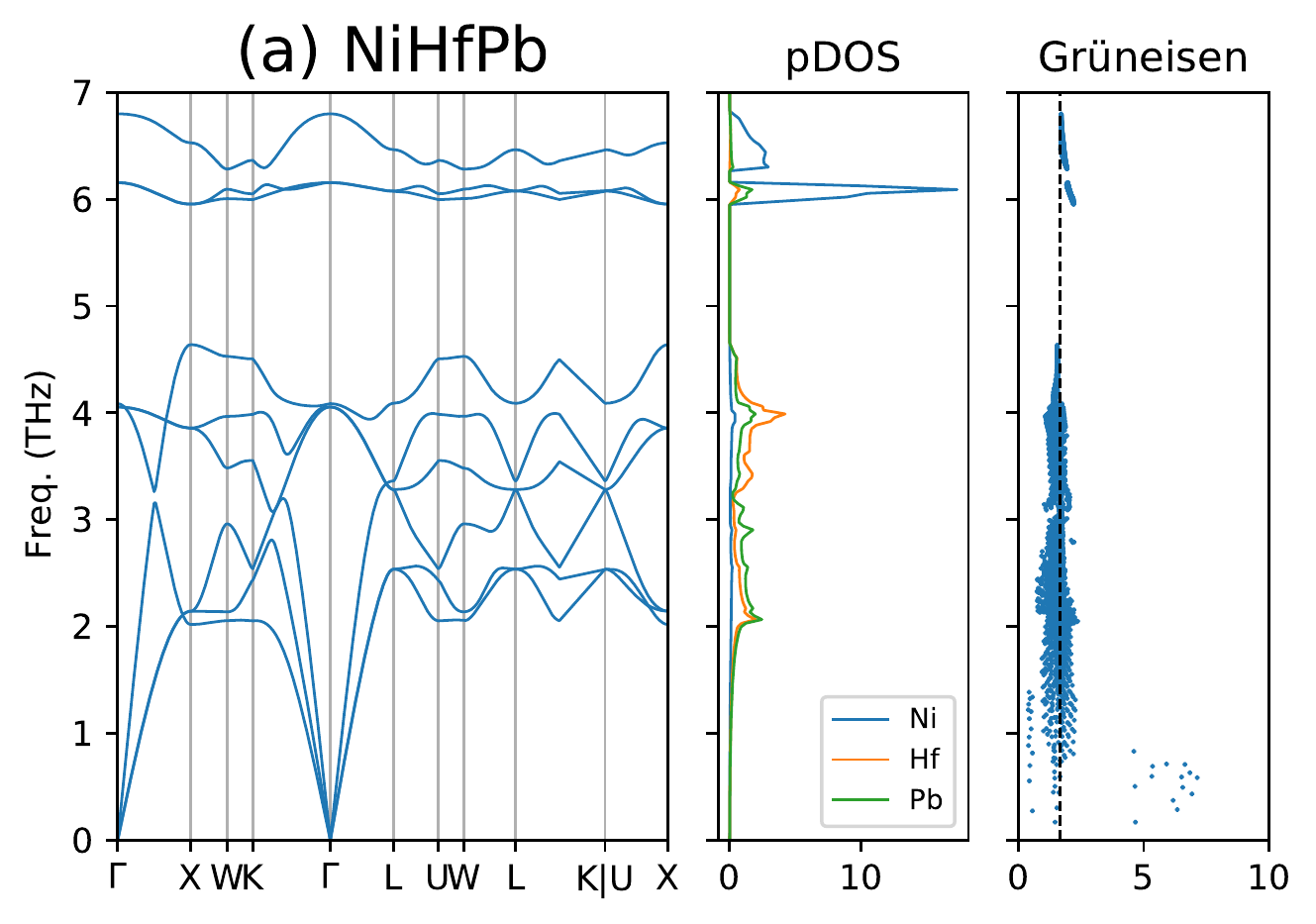}
    \includegraphics[width=0.6\columnwidth]{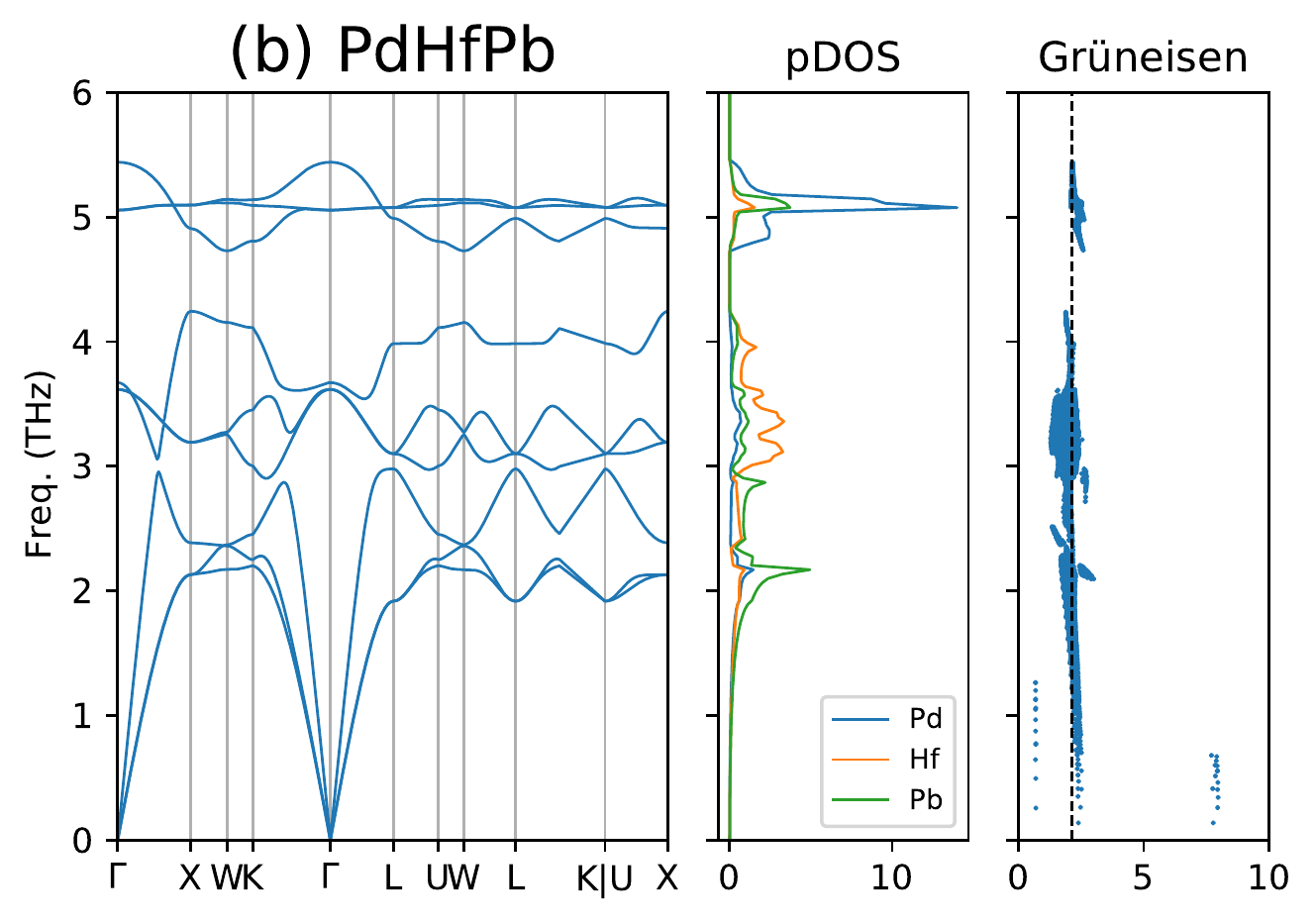}
    \includegraphics[width=0.6\columnwidth]{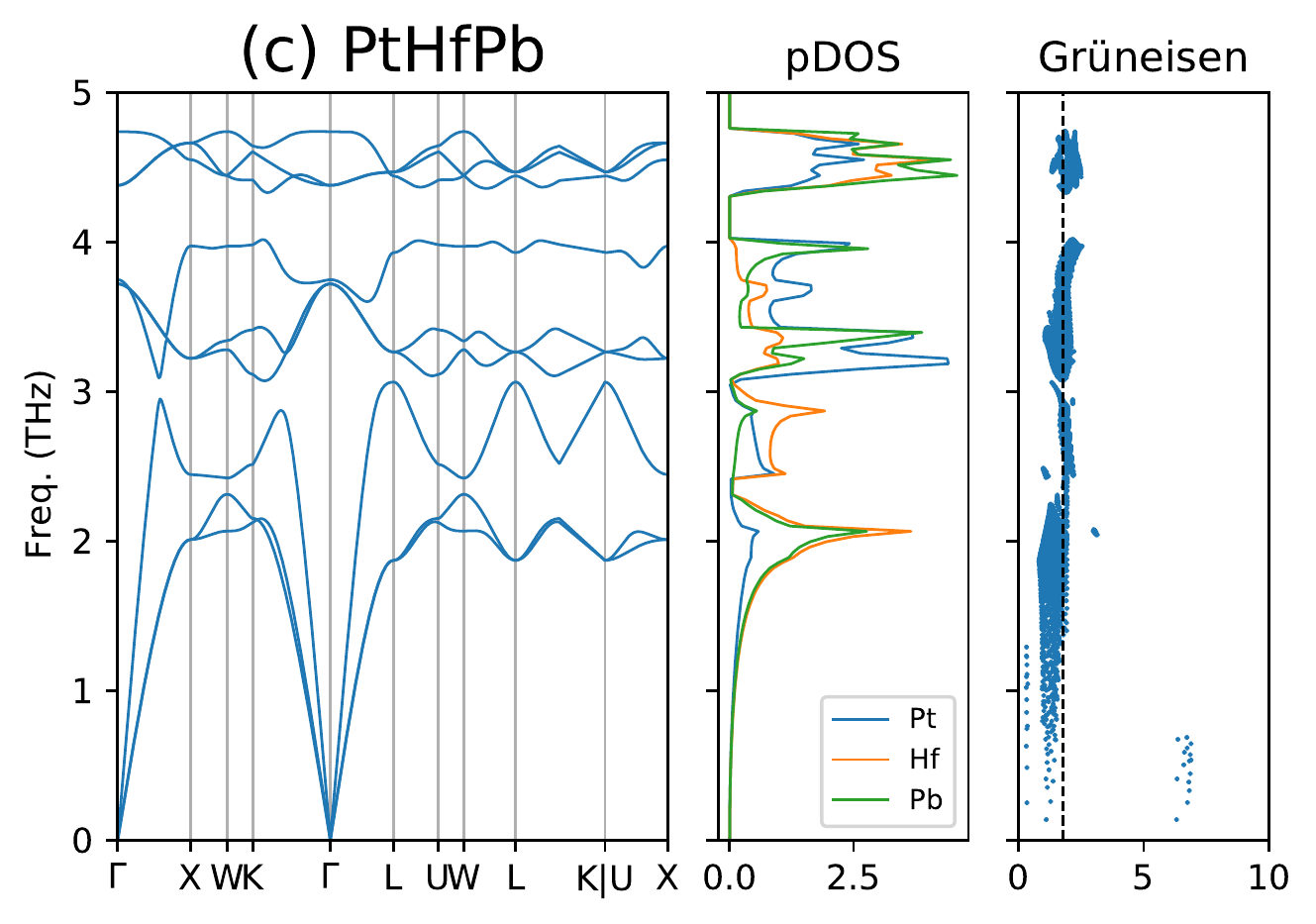}
    \caption{
       Phonon dispersion, atom-projected phonon density of states and scatter plot of the mode Gr\"uneisen values of (a) NiHfPb, (b) PdHfPb and (c) PtHfPb. The vertical lane in the rightmost panes marks the averaged Gr\"uneisen parameter.}
    \label{fig:phonon}
\end{figure}

Next, we calculated the phonon dispersion of the three alloys. We first established the absence of any negative frequency in all these materials, confirming their dynamical stability. In each compound, there are three atoms in their primitive unit cell, resulting in 6 upper (optical) and  3 lower (acoustic) modes, as shown in Fig.~\ref{fig:phonon}.

Since the alloys are semiconductors we first computed the high frequency dielectric constant $(\epsilon_\infty)$ and Born effective charges ($Z^\star$) using Density Functional Perturbation Theory~\cite{Gonze1995a,Gonze1995b,Baroni2001}. The partial ionic character of these alloys gives rise to the LO-TO splitting of the high frequency optical modes. The results are reported in Tab.~\ref{tab:dielectric}. The dielectric constant of NiHfPb and PdHfPb is $\sim$23 since the two alloys have a similar band gap. The dielectric constant of PtHfPb is slightly smaller, due to the larger band gap. During atomic vibrations, the X atom drags most of the electrons, depleting Hf and Pb, which behave as positively charged. The large electron redistribution is an indication of strong hybridization between the orbitals of X with the other two atoms. Therefore the chemical bond can be classified as covalent with a partial ionic character.

\begin{table}[tb!]
    \centering
    \begin{tabular}{c c c c c c c}
    Alloy & $\epsilon_\infty$ & \multicolumn{5}{c}{Born charges $Z^\star$}  \\
             &                   & Ni & Pd & Pt & Hf & Pb \\
    \hline
    NiHfPb & $23.003$ & $-3.630$ & & & $2.686$ & $0.944$ \\
    PdHfPb & $23.505$ & & $-3.505$ & & $2.745$ & $0.760$ \\
    PtHfPb & $21.522$ & & & $-3.818$ & $2.361$ & $1.457$ \\
    \end{tabular}
    \caption{Calculated high frequency dielectric constant ($\epsilon_\infty$) and Born charges ($Z^\star$).}
    \label{tab:dielectric}
\end{table}

For the three alloys, the acoustic (optical) modes are located below (above) $\sim$3.2\,THz. For NiHfPb, the highest optical modes are situated in the region above 6\,THz and are mainly contributed by the Ni atom. Similarly for PdHfPb, the optical modes are at about 5\,THz and are localized mainly on the Pd atom. The reason for such separation of the modes is the large difference between the mass of Ni (58.69), Pd (106.42) and the other atoms. The situation is different in PtHfPb since the mass of Pt (195.08) is similar to the mass of Hf (178.49) and Pb (207.20). Here all thee atoms contribute equally to the optical modes and no clear separation of atomic motion can be observed. In the three alloys the acoustic modes are a superposition of Pb and Hf vibrations.

In Fig.~\ref{fig:phonon} we also report the scatter plot of the mode Gr\"uneisen parameters, as a function of the phonon frequency. These have been calculated with PHONOPY on a dense mesh of 20$\times$20$\times$20 points in reciprocal space. Then we averaged the Gr\"uneisen parameters over the Brillouin zone and we reported the average as a vertical line in Fig.~\ref{fig:phonon}, as well as the numerical value in Table~\ref{tab:gruneisen}. In the three alloys considered the Gr\"uneisen parameters are distributed narrowly around their average value. The optical modes of NiHfPb and PdHfPb display Gr\"uneisen values larger than the average. The exception is the acoustic modes below 1\,THz that show a large spread in the Gr\"uneisen values. Such large spread is significant for the modified Debye-Callaway model of thermal conductivity. Overall NiHfPb and PtHfPb have similar average Gr\"uneisen, while it is larger for PdHfPb, indicating a more pronounced anharmonic degree in the latter system.

\begin{table}[tb!]
    \centering
        \begin{tabular}{c c}
        Alloy & Average Gr\"uneisen  \\
        \hline
        NiHfPb & 1.658  \\
        PdHfPb & 2.116  \\
        PtHfPb & 1.772  \\
        \end{tabular}
    \caption{Average Gr\"uneisen parameter, averaged over the Brillouin zone.}
    \label{tab:gruneisen}
\end{table}

\subsection{Thermoelectric properties}
\begin{figure}
    \centering
    \includegraphics[width=0.6\columnwidth]{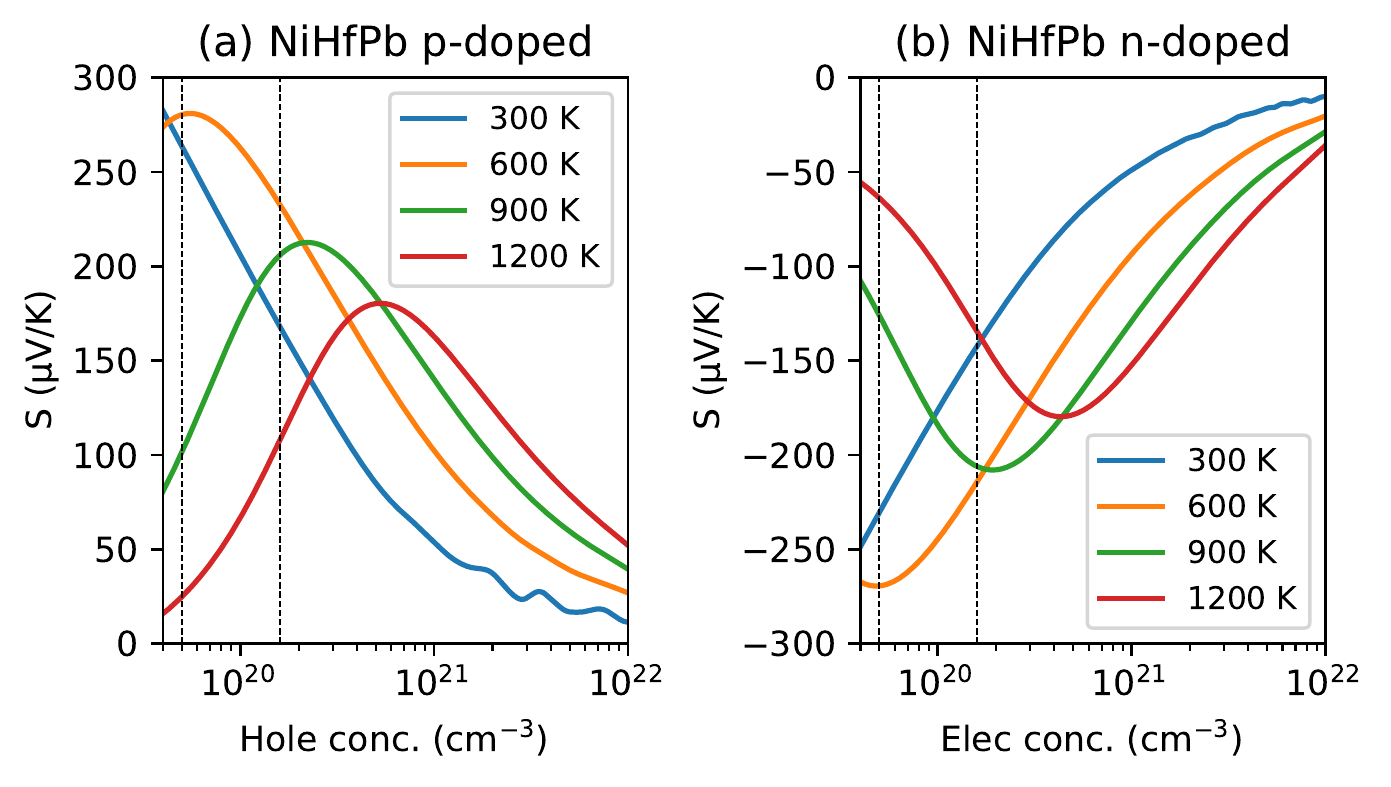}
    \includegraphics[width=0.6\columnwidth]{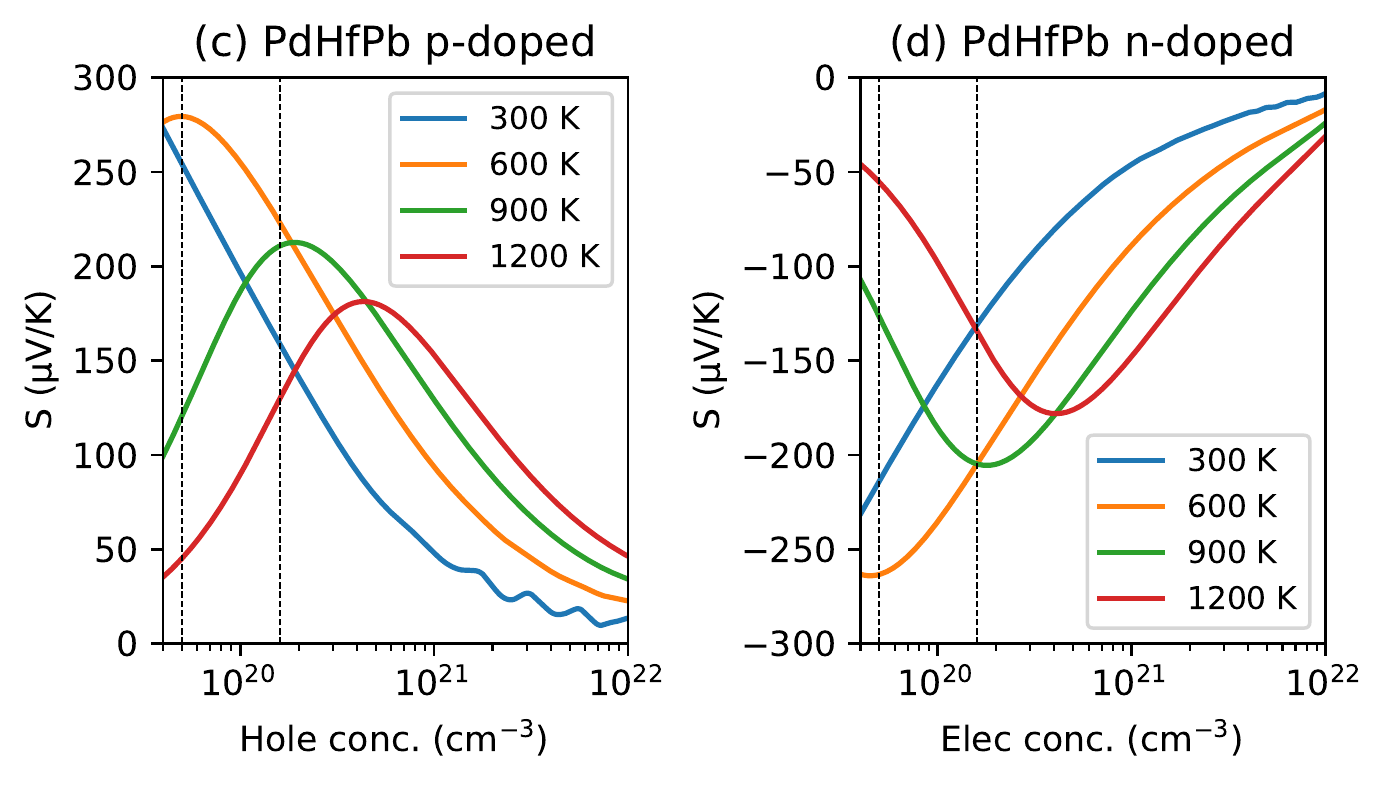}
    \includegraphics[width=0.6\columnwidth]{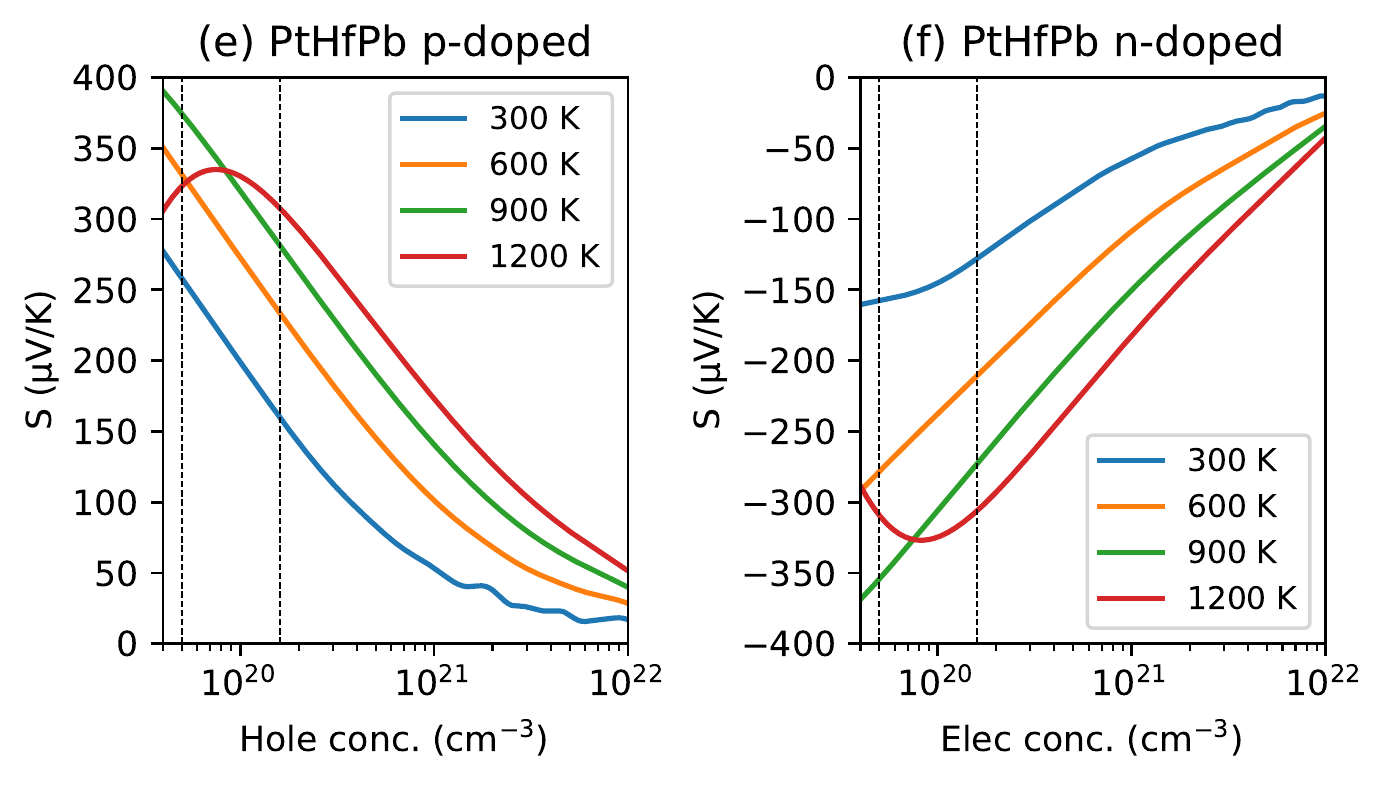}
    \caption{Temperature dependence of the calculated Seebeck coefficient as a function of the hole and electron concentration for (a,b) NiHfPb, (c,d) PdHfPb and (e,f) PtHfPb.}
    \label{fig:seebeck}
\end{figure}

The trend of the Seebeck coefficient ($S$) as a function of the carrier concentration $n$ (electrons or holes) at different temperatures $T$ of the alloys is reported in Fig.~\ref{fig:seebeck}. We have hypothesized two doping regimes, marked by the vertical dashed lines: ``low'' doping, corresponding to $|n|=5.0\cdot10^{19}\,\text{cm}^{-3}$ and ``high'' doping, corresponding to $|n|=1.6\cdot10^{20}\,\text{cm}^{-3}$. Note that the high doping regime is equivalent to an excess charge of $\sim$0.01 electrons/holes per unit cell. However, using Boltztrap, the doping is simulated in the rigid band approximation, i.e. without adding or subtracting any fraction of charge in the DFT calculation.

The first observation is that NiHfPb and PdHfPb behave very similarly. In the case of low-doping with hole carriers, the highest value of the Seebeck coefficient is $S\sim 300$\,$\mu$V/K, obtained at the lowest temperature, $T=300$\,K. As the hole concentration increases (Fig.~\ref{fig:seebeck}a,c) the Seebeck coefficient reaches a maximum value and then it decreases. As the temperature is increased, the position of the $S$ peak shifts towards larger carrier densities, and at the same time the height of the peak decreases. A similar behavior can be found for electron doping (Fig.~\ref{fig:seebeck}b,d). Here, the Seebeck coefficient is negative and reaches the largest absolute value of $\sim 250$\,$\mu$V/K at low-doping and low temperature. The similarity between NiHfPb and PdHfPb is not unexpected, given the similar shape of the density of states and the similar band gap. Therefore, for these two alloys we expect $S$ to decrease monotonically with $T$ for any given carrier concentration.

The situation is remarkably different in the PtHfPb alloy (Fig.~\ref{fig:seebeck}e,f). Here the maximum value of the Seebeck coefficient can be obtained for very low carrier doping up to $T=900$\,K. The Seebeck coefficient is predicted to increase with temperature both at low and high doping. Moreover, the maximum value of $|S|$ is roughly twice as much with respect to the other two alloys. The reason of this peculiar behavior of PtHfPb is rooted in the fine details of the band structure. The bowing of the first conduction band along $\Gamma-X$ is probably causing a different mixing of the wavefunctions both of the VBM and the CBM, with an enhanced contribution from Pt and Pb.

\begin{figure}
    \centering
    \includegraphics[width=\columnwidth]{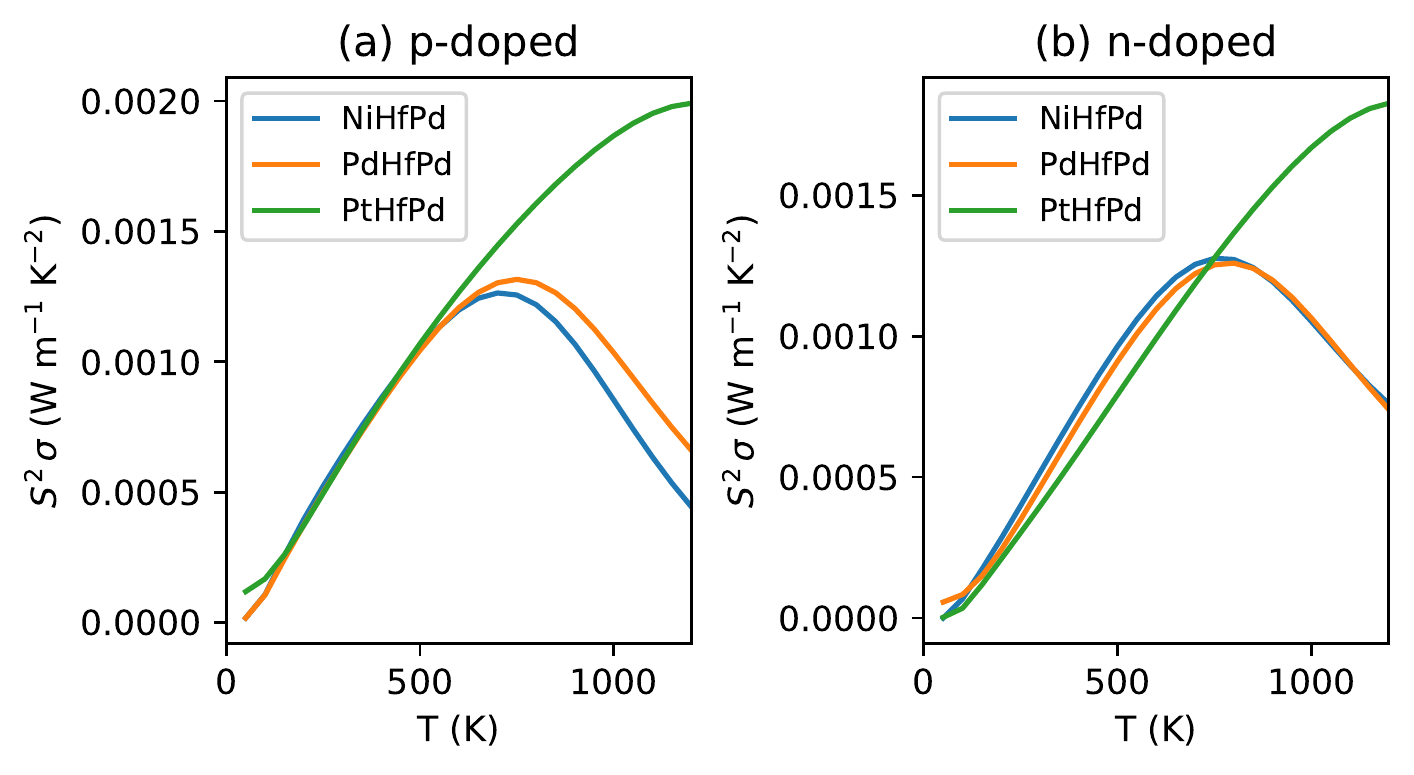}
    \caption{Temperature dependence of the power factor $S^2\sigma$ calculated with $\tau=3.18\cdot10^{-15}$~s and at ``high'' doping $|n|=1.6\cdot10^{20}\,\text{cm}^3$.}
    \label{fig:powerfactor}
\end{figure}

To calculate the power factor as a function of carrier density and temperature, one would need to estimate the relaxation time $\tau$. In absence of experiments on these alloys, we use the experimental value deducted for ZrHfPb~\cite{Gautier2015,Wang2016}, i.e. we assume a constant relaxation time $\tau$ of $3.18\cdot10^{-15}$~s at room temperature and carrier density of $1.6\cdot10^{20}\,\text{cm}^{-3}$. With this values, the power factor is show in Fig.~\ref{fig:powerfactor}. The power factor of the three alloys is similar from low temperatures until 500~K. However, above 500~K the power factor of NiHfPb and PdHfPb quickly degrades, whereas the power factor of PtHfPb increases monotonically with temperature.

\begin{figure}
    \centering
    \includegraphics[width=\columnwidth]{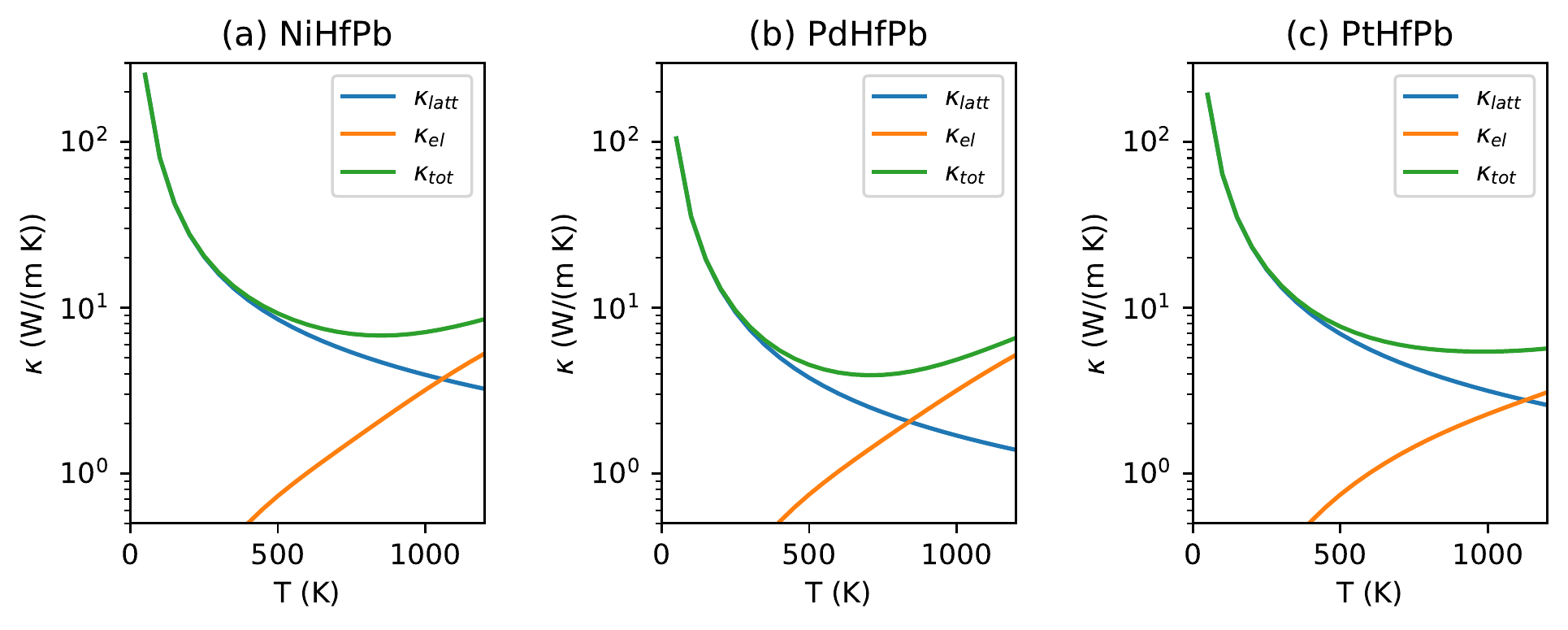}
    \caption{Total thermal conductivity as a function of temperature for NiHfPb, PdHfPb and PtHfPb. The lattice thermal conductivity is  with the modified Debye-Callaway model~\cite{Morelli2002,Fan2020,Fan2021}. The electronic thermal conductivity is calculated within the CRTA, for the ``high'' doping concentration of holes}.
    \label{fig:kappa_tot}
\end{figure}

Next, in Fig.~\ref{fig:kappa_tot} we report the calculated thermal conductivity  ($\kappa=\kappa_{latt}+\kappa_{el})$ as a function of temperature for the three alloys. In all the cases $\kappa_{latt}$ decreases at larger $T$ due to growing anharmonic effects. NiHfPb has the largest thermal conductivity, followed by PtHfPb. Among the three materials, PdHfPb has the lowest thermal conductivity. According to Tab.~\ref{tab:gruneisen} NiHfPb has the lowest average Gr\"uneisen parameter (i.e. the lowest anharmonicity). On the other extreme, PdHfPb has the largest Gr\"uneisen parameter (i.e. the highest anharmonicity). At room temperature, the calculated lattice thermal conductivity is found to be $7.3$, $13.2$, and $16.0$~W~m$^{-1}$K$^{-1}$ for PdHfPb, NiHfPb and PtHfPb, respectively. Our calculated small lattice thermal conductivity of PtHfPb is in agreement with previous calculations from Kaur et al.~\cite{Kaur2017}.

{\color{red} On the contrary the electronic contribution to the thermal conductivity $\kappa_{el}$
increases with temperature, up to 5 W/(m K) at 1200~K. This is shown in Fig.~\ref{fig:kappa_tot} in the high doping regime in the CRTA. Note that the $\kappa_{el}$ does not change appreciably upon hole or electron doping.

At low temperature $\kappa$ is dominated by the lattice contribution up until $\sim$600~K, where the electronic term compensate the lattice term and the total thermal conductivity remains nearly constant with temperature. This behavior shows that the electronic term cannot be neglected for high temperature Heulser-based thermoelectric generators.}

\begin{figure}
    \centering
    \includegraphics[width=\columnwidth]{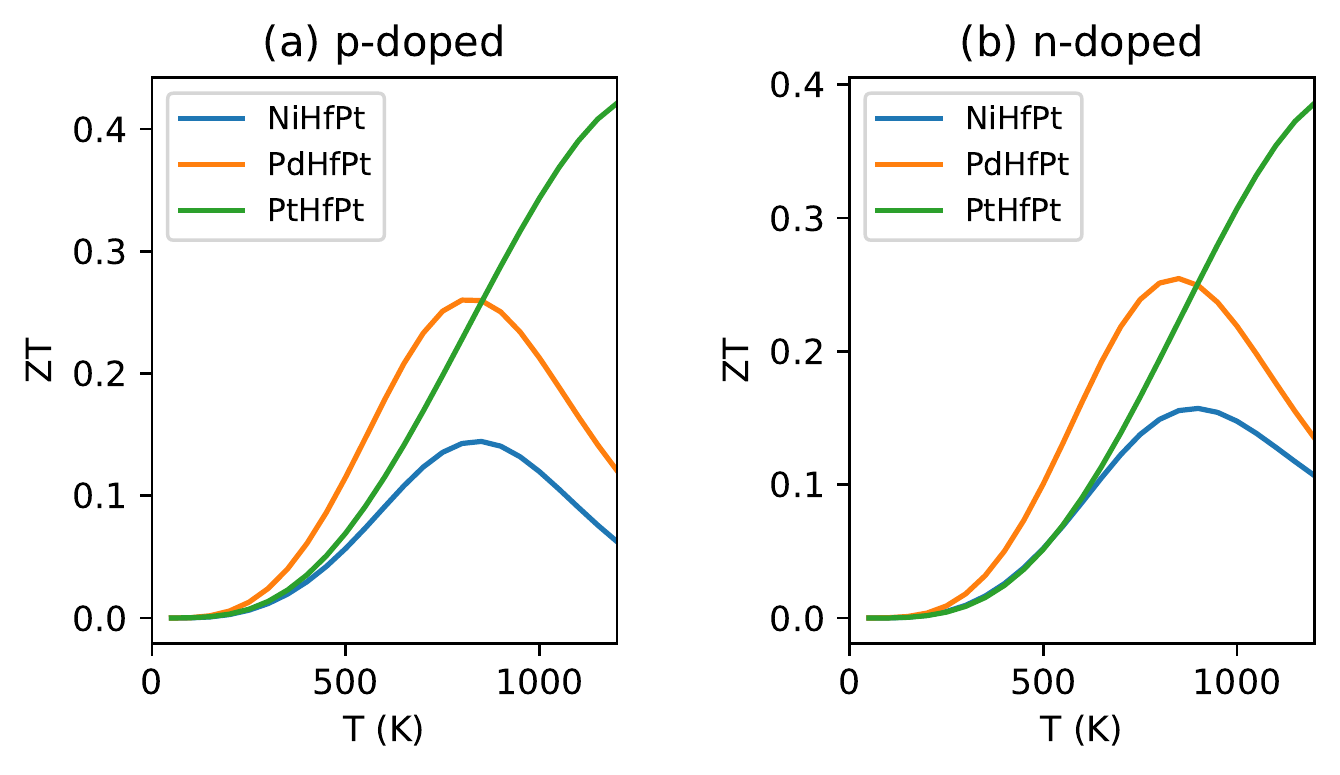}
    \caption{Figure of merit as a function of temperature for NiHfPb, PdHfPb and PtHfPb for $\tau=3.18\cdot10^{-15}$\,s and high doping regime.
    }\label{fig:ZT_CRTA}
\end{figure}

With all the above ingredients we can finally evaluate the figure of merit $ZT$, which is reported in Fig.~\ref{fig:ZT_CRTA}. Similarly to the power factor (Fig.~\ref{fig:powerfactor}) the figure of merit of the three alloys increases with temperatures up to $\sim$~900~K. Then, at higher temperature the $ZT$ of NiHfPb and PdHfPb tends to decrease. This is not so for PtHfPb, whose figure of merit reaches a respectable value of 0.4 at 1200~K, and it would further increase. {\color{red} This is particularly relevant since the typical operating temperature of a high-T thermoelectric generator is in the 1200-1400~K range}. Note that our PtHfPb figure of merit is larger than that of 0.25 found by Kaur et al.~\cite{Kaur2017}. The discrepancy might be due to the a different method to calculate the lattice thermal conductivity, but, most importantly, because they used a different crystal structure, which is not the most thermodynamically stable.

\section{Conclusions}\label{sec:conclusion}
We have investigated in detail the thermoelectric properties of three newly discovered half-Heusler alloys XHfPb (X = Ni, Pd, and Pt) by combining the first-principles density functional theory (DFT), semi-classical Boltzmann transport theory, dynamical matrix calculation, as well as the modified Debye-Callaway model. We calculated both the electronic and lattice contribution to figure of merit, with experimental values for the carrier concentration $|n|$ and for the relaxation time $\tau$, in order to make realistic prediction.
The electronic properties of three alloys appears to be similar. Their power factor increases monotonically with temperature up to 500~K. However, at high temperatures, PtHfPb behaves differently from NiHfPb and PdHfPb. In fact, the power factor of PtHfPb keeps increasing with temperature in a large range of carrier doping. As a results, at high doping PtHfPb shows a respectable figure of merit of $\sim$0.4. Therefore we conclude that among the three alloys PtHfPb is better suited for high temperature thermoelectric applications, even at lower carrier densities. However the thermal conductivity of PtHfPb is intermediate between the other two alloys, and mixing of Pt and Pd on the X site of the half-Heusler structure might be beneficial to further enhance the figure of merit.

\section*{Acknowledgments}
The Authors wish to thank Alberto De Bernardi for useful discussion. The work has been performed under the Project HPC-EUROPA3 (INFRAIA-2016-1-730897), with the support of the EC Research Innovation Action under the H2020 Programme. In particular, POA and GAA gratefully acknowledge the support of DC and RG for hosting them. Computer resources and technical support provided by CINECA are also acknowledged.



\end{document}